\documentclass{article}

\usepackage{graphicx}
\usepackage{amssymb}

\usepackage{boxedminipage} 

\begin{document}
\title{A class of structured P2P systems\\ supporting browsing}

\author{Julien Cohen\footnote{Email address: \texttt{Firstname.Name@univ-nantes.fr}}\\ LINA\footnote{UMR 6241 Universit\'e de Nantes, \'Ecole des Mines de Nantes, CNRS}\\
 Universit\'e de Nantes (Polytech'Nantes)}

\date{July, 2009}

\maketitle

\begin{abstract}
Browsing is a way of finding documents in a large amount of
data which is complementary to querying and which is
particularly suitable for multimedia documents.
Locating particular documents in a very large
collection of multimedia documents such as the ones
available in peer to peer networks is a difficult
task. However, current peer to peer systems do not allow to
do this by browsing.

In this report, we show how one can build a peer to peer
system supporting a kind of browsing. In our
proposal, one must extend an existing distributed hash table
system with a few features : handling partial hash-keys
and providing appropriate routing mechanisms for these
hash-keys.
We give such an algorithm for the particular case of the
Tapestry distributed hash table.

This is a work in progress as no proper validation has been done
yet.

\end{abstract}

\section{Introduction}

Browsing is a means to access information which is
complementary to querying. A browsing system is useful if
it allows the user to find quickly the document she is
looking for among a very large set of documents. 
In current peer to peer (P2P) systems, browsing is not
supported and the user has to make with keyword
queries.
This kind of queries is convenient for textual
data but not for multimedia documents.

In this paper, we show that one can easily extend
distributed hash tables (DHTs), which are simple and efficient P2P
systems, to support browsing.
To illustrate this, we build a personal image collection
sharing application supporting browsing. We also show with
an example how the routing mechanisms of DHTs can be
extended to ensure that browsing makes a convenient use of the
network.

\subsection*{Browsing and peer to peer}

Characteristics of fully decentralized peer to peer systems
(without any kind of server or super-peer) make browsing
difficult to achieve.
Unstructured systems like Gnutella are not quick enough to
answer queries due to the flooding method for locating
data. Indeed, browsing is an interactive process between the
user and the system and needs short latency.
On the opposite, structured systems (DHTs) are very fast but
the query mechanism is too poor to deal with document
properties.
Nevertheless, we will show how to extend that query mechanism while
preserving most of the good properties of DHTs.

\subsection*{Providing complex queries in DHTs}

In DHTs, document identifiers are hashed by classical hash functions
and routing mechanisms make it possible to
associate a unique machine in the system to each
hash key. Hash keys then play the role of logical addresses.
In that setting, one must know the exact identifier of the
object she is looking for to compute its hash key in order
to locate it in the system. This clearly does not allow
search by content.

To solve this problem, we use combinatorial hashing~\cite{knuth3}. Let us
explain this. First, instead of hashing an identifier, we
hash a description, \emph{i.e.} a set of properties, of the documents. This means that all the
documents with the same properties have the same
hash key. One can then ask to locate all the documents with
some desired properties since she can compute the
corresponding hash key.
Second, instead of hashing the whole property descriptor, we
hash each property separately into small hash keys and we
concatenate them into a large hash key. Then, hash keys being
structured into sub-hash keys, we can replace some slices with
wildcards to handle all hash keys with a given value for some
properties of interest.

The first of these two points is theoretically sufficient to
allow search by content since hash keys with wildcards can
be compiled into a set of simple hash keys.
However, that solution is not convenient in practice because
in general the corresponding set of descriptors is very
large and it would lead to flooding the network.
This is why structuring hash keys is useful as it will allow
to make a convenient use of the network by the means of
relevant locating mechanisms.


\section{Descriptors, hash keys, wildcards}

\subsection{Descriptors}

We assume documents are described by vectors of properties
called \emph{descriptors}. 
All the descriptors in the system must have the same
structure : the same number of components, component at a same position
representing a same property, with a shared meaning for the possible
values of a component.
Descriptors are usually extracted automatically even if it
can be done manually. We accept this extraction to be costly
since it is done on the machine which is sharing the document
(it is fully decentralized).

One has to keep in mind the goal of browsing when choosing
the descriptor space. Indeed, we use descriptors to denote
classes of documents among which the browsing takes place.
First, the properties descriptors hold must be meaningful
for the user. For instance, for personal image collections,
users usually do not make a distinction between GIF and
JPEG images. 
Second, the range of values for each property must be coarse
grained.  For instance, if we consider the brightness of an
image, 256 possible values is too much. A choice of 3 values
(dark, medium, bright) could be better for browsing.
Of course, the choice of the descriptor space largely depends on
the intended application and on the available technologies
for the analyze of the considered type of documents.

In the following, we call \emph{digits} the components of
descriptors and we call \emph{range} of a digit the set of
possible values for the corresponding property.

Let us give two examples of descriptor spaces for an
application of sharing personal image collections.

\begin{figure}
\begin{center}
\includegraphics[width=4cm]{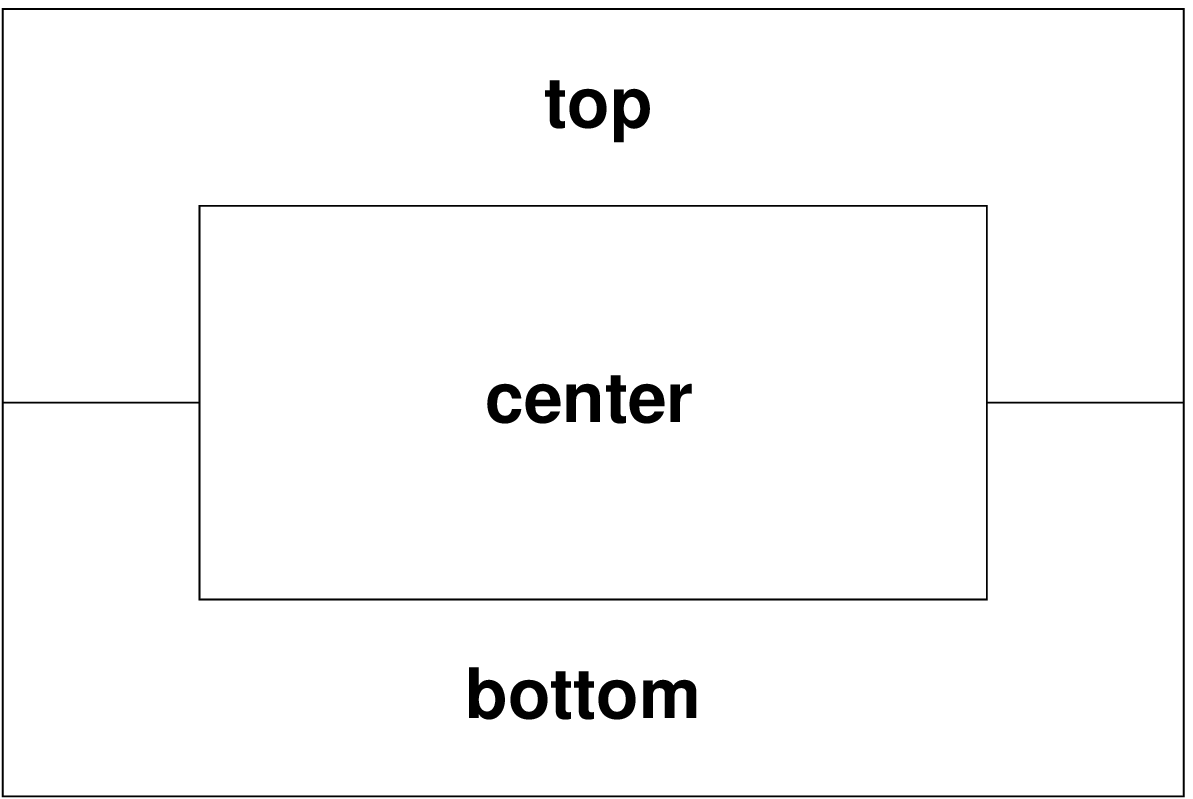}
\end{center}
\medskip

\hfill
\includegraphics[width=4cm]{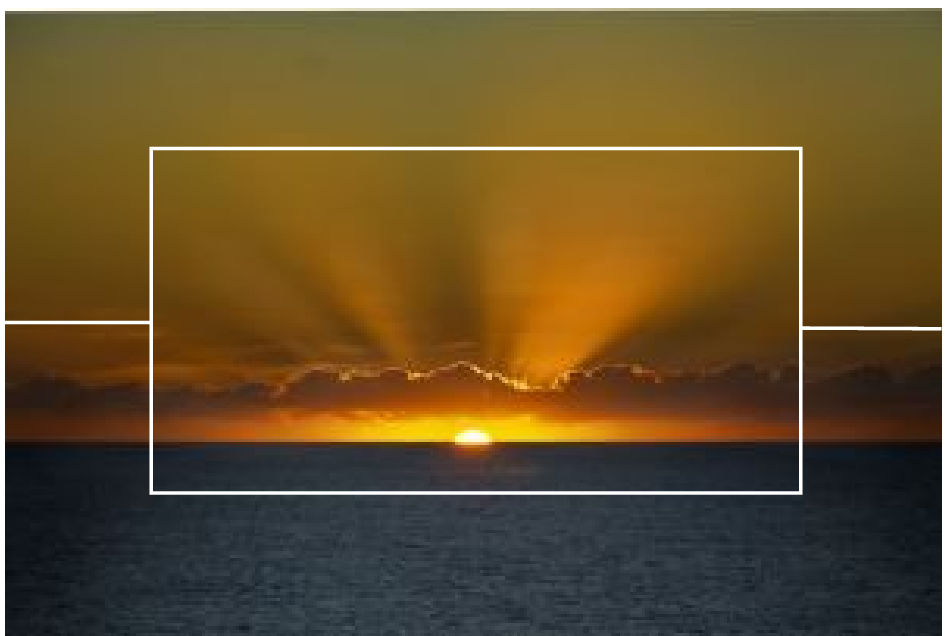}
\hfill
\includegraphics[width=4cm]{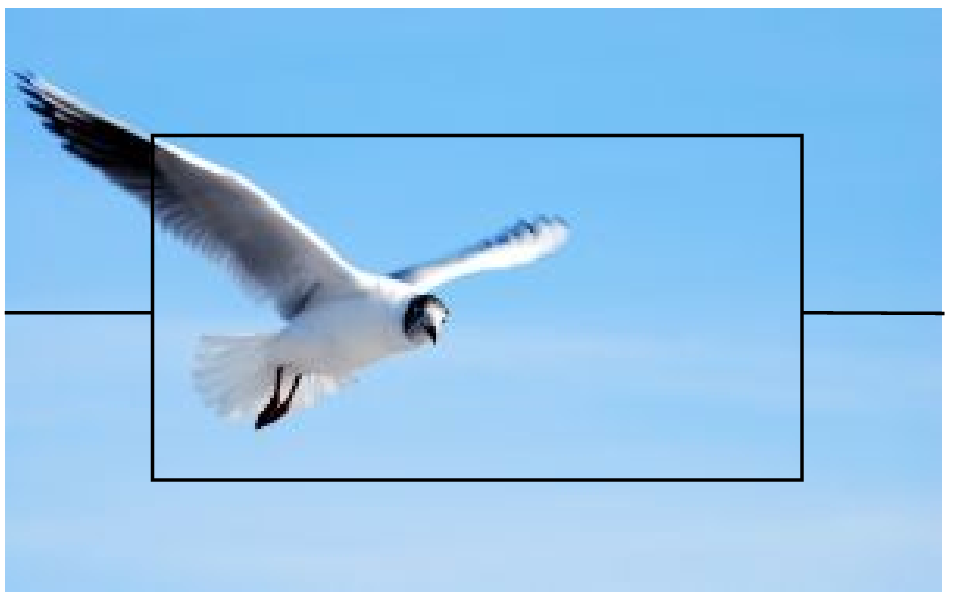}
\hfill\phantom.

\caption{Example of division of an image into areas}
\label{fig-decoupage}
\label{fig-photos}
\end{figure}

\subsubsection{Toy example.}

The considered properties  are the brightness in
several characteristic areas of the picture. For each area,
the mean brightness is normalized into the $\{0,1\}$ space.
The areas are the center, the bottom and the top area, as
pictured in figure~\ref{fig-decoupage}.  This choice of
areas is justified by the fact that on many pictures, the
subject is near the center and is surrounded by its
environment.
Here a descriptor is then a sequence of 3 binary digits, 000 meaning
a dark picture and 111 meaning a bright picture.

\subsubsection{Second example.}
In this second example, descriptors are composed of average
values for red, green, and blue components of the same three
areas of the picture.
Descriptors then have 9 digits.
In addition, these values are normalized into integers
between 0 and 3.
With this setting, digits are encoded onto 2 bits and
descriptors onto 18 bits.

If we consider the sunset picture of
figure~\ref{fig-photos}, the descriptor is 000220320,
that is 0, 0, 0 for the bottom area (black), 2, 2, 0 for the
middle area (yellow), and 3, 2, 0 for the top area
(orange).
For the flamingo picture, the
descriptor is 223322223, that is 2, 2, 3 for the bottom area
(light blue), 3, 2, 3 for the middle area (light red and blue) and 2, 2, 3
for the top area.



\subsection{Hash keys}

In distributed hash tables, hash keys are used as logical
addresses in the system. These addresses are sequences of
bits of a chosen length. 
Descriptors have to be transformed into hash keys. (We could
also have extended a DHT to accept descriptors as logical
addresses which gives approximately the same result).
For each digit of descriptors, if $n$ is the size of the
range, we set a bijection from that range to the set of
integers between 0 and $n-1$. These bijections are used as
``hash'' functions for the digits of the descriptors.
A hash key is then the sequence of the hashed digits.



Mapping descriptors to hash keys allows to map each
descriptor to a logical address of the system (we take as
set of logical addresses the set of possible hash keys).
Then, to locate all the documents having a given descriptor,
one just has to send a message to the corresponding logical
address (in DHTs, the system allows to deliver a message to
a machine knowing only its logical address).
Since there is a bijection between descriptors and hash keys,
we will consider in the rest of this paper that these notions are the same.

Note that in Tapestry~\cite{tapestry2001}, the DHT we use as
an example, the digits of the tuples are originally in a same base,
which is a power of 2, but this can be easily relaxed to any
range of integers (or to any finite set equipped with a
total order), which does not need to be the same for all the
digits of the tuples.
So the ranges of the different digits of the tuple can
be chosen independently to describe different sorts of
properties of the documents.

\subsubsection{Discussion.}

Unlike traditional hash keys, descriptors of available documents are
not uniformly distributed in the key space. This could break
the balance of the system since a small number of machines
could be responsible for the location of popular types
of documents.  Depending on the particular application, this
could be a major problem.
However, we can notice that :
\begin{itemize}

\item
Each node is responsible for several descriptors, possibly a
large number (depending on the number of participants and
on the size of the descriptor space). So each machine could be
responsible of popular and unpopular descriptors.

\item
Location queries are light requests compared to the cost of
transferring a document and the documents are not
located on the node which is responsible of their location.

\item
DHTs usually use a cache of location to allow responsible nodes to be helped by their
logical neighbors to answer location queries.
\end{itemize}

These remarks make us think that in most cases the unbalance
of the descriptors is negligible.

\subsection{Descriptors with wildcards, hash keys with wildcards}

We use $*$ wildcards (read \emph{star wildcards}) instead of
some digits in descriptors to denote a property which value
has not been fixed.
Descriptors with wildcards can be seen from two points of view.
First, it can denote a set of descriptors. For instance,
0*0 denotes the set $\{000, 010\}$.
Second, it can denote the values of some properties, in
particular to represent what kind of document the user is
looking for. For instance, 0*0 represent the properties
``dark bottom'' and ``dark top'' in our toy example.
We will see that the browsing process builds a descriptors
with wildcards that expresses users query, but the user does
not need to see that descriptor.
In the rest of the paper, ``descriptor with wildcards'' means ``descriptor that can have 
wildcards''.

We say that a descriptor is \emph{denoted} by a 
descriptor with wildcards $d$ if it can be obtained
by replacing the wildcards of $d$ by appropriate values.

As for descriptors, we will use hash keys with wildcards
(this notions depends on the type of DHT, for Tapestry, a
wildcard replaces a digit). Wildcards in descriptors are
mapped to wildcards in hash keys. 

\section{Browsing, samples}
\label{sec-browsing}
\subsection{The browsing process}
From the user point of view, the browsing process is based
on the following loop:
\begin{enumerate}

\item \label{display-step} 
The system displays the representation of some documents satisfying the expressed constraints.

\item Based on that display, if the user is satisfied,
then she asks for previewing the corresponding documents,
else she indicates to the system an additional
constraint to take into account\footnote{
In order to provide a better browsing process, one should
also consider the possibility of removing one of the
constraints already set. However, we do not consider this in
this report.
}.

\end{enumerate}

Of course, the hard work is hidden from the user. 
From the system point of view, the browsing process roughly
consists in building iteratively  a
descriptor (possibly with wildcards) expressing user's needs.
The descriptor with wildcards under construction, called the
\emph{current descriptor}, evolves at each browsing step.
The process starts with a current descriptor containing only $*$
wildcards and progresses by replacing wildcards by values.
Each step in the process consists in:
\begin{enumerate}

\item 
Retrieving the representation of some documents which are
representative of the documents which are denoted by the
current descriptor.

\item Displaying them.

\item Allowing the user express a preference on the displayed documents.

\item Deciding based on that input 
 which wildcard to replace with which value.

\item Making the corresponding change in the current descriptor
\end{enumerate}

The loop ends when the current descriptor has not any wildcard
left, or when the user is satisfied with the current result.

The visual representation of documents is defined is
section~\ref{sec-miniature}. 
A notion of representativeness is proposed in section~\ref{sec-samples}.
The strategy for retrieving the locations of representative
documents given a descriptor with wildcards is described in
section~\ref{sec-sample-resolution}.
A simple way for making the user choose the evolution of the
current descriptor is proposed in
section~\ref{sec-interface}.
The retrieval of all the documents at the end of the process
is discussed in section~\ref{sec-total-resolution}.

\subsection{Miniatures}
\label{sec-miniature}

During the browsing process, it is not necessary to have a
whole document to show to the user, a short description
allowing her to preview it and which is cheaper to transfer is
sufficient. We call this short version a \emph{miniature}.
Miniatures can be thumbnails for pictures, storyboards for
videos, summaries for texts, extracts for audio and so on.
During the browsing process, we only use
miniatures. A machine which publishes a document must also
provide its miniature.

\subsection{The browsing graph}

\begin{figure}
\begin{center}
\includegraphics[width=10cm]{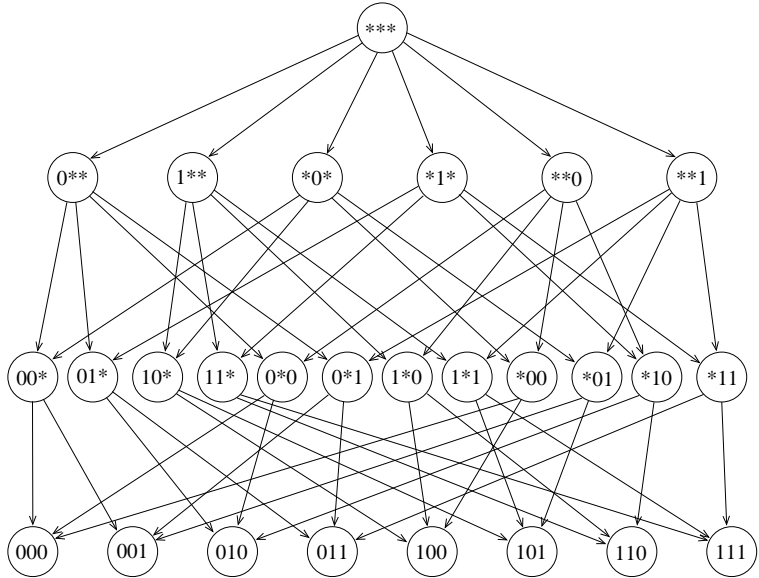}
\end{center}
\caption{Browsing graph}
\label{fig-treillis}
\end{figure}

The browsing process can be described by a finite state
machine where each state corresponds to a descriptor with
wildcards (the current descriptor). The state machine for
our toy example is pictured in figure~\ref{fig-treillis}. We
call this state machine the browsing graph. The browsing
graph, which is a lattice, only depends on the descriptor space.

\subsection{Illustrative documents, samples, representative sets}
\label{sec-samples}

Displaying one document for each descriptor denoted by a
descriptor with wildcards would make too much information to
retrieve, to display and to handle for the user. 
We make the assumption that  displaying only one document for each
direct successor of the current descriptor in the browsing
graph is sufficient. Let us formalize this idea.

\subsubsection{Illustrative documents.}
We say a document is \emph{illustrative} for a descriptor
with wildcards $d$ when its descriptor is denoted by $d$.

\subsubsection{Samples.}
Given a descriptor with wildcards $d$, we call \emph{sample} a set of
documents that contains
for each descriptor with wildcards which is a
direct successor of $d$ at least one illustrative document
and which does not contain any document which is not
illustrative for $d$.

\subsubsection{Illustrativeness and the browsing graph.} We have the following
property: a document is illustrative for an descriptor with
wildcards $d$ if and only if the descriptor of that document
is reachable from $d$ in the browsing graph.
This shows that displaying  a sample is sufficient
to allow the user to reach any descriptor that is denoted by
the current descriptor by browsing. 

%
%

\subsubsection{Representative sets of descriptors.}
Finally, we introduce the notion of representative set of
descriptors which is used in our routing algorithm to
retrieve samples.
Given an descriptor with wildcards $d$,
a set $S$ of descriptors is said to be \emph{representative}
when ($d[i]$ denotes the digit of $d$ in position
$i$):\begin{itemize}

\item
for each $i$, if $d[i] \not = *$, then for each descriptor
$e$ in $S$ we have $e[i]=d[i]$,

\item
 for each $i$ such that $d[i] = *$ and for
 each value $v$ in the range of the digit in position $i$,
 there is a descriptor $e$ in $S$ such that $e[i] = v$.

\end{itemize}

Having one document for each descriptor of a representative
set of descriptors (for $d$) constitutes a sample (for $d$).

\subsection{User interface}
\label{sec-interface}

\begin{figure}[ht]
\begin{center}
\includegraphics[height=3.8cm]{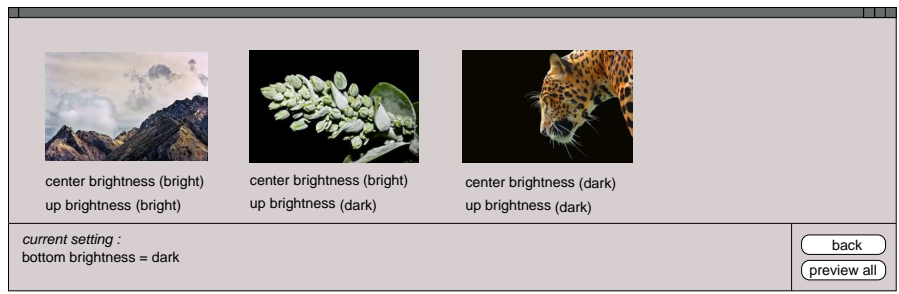}
\caption{Example of a possible browsing interface}
\label{fig-interface}
\end{center}
\end{figure}

%
A document with descriptor $d'$ is illustrative for
all the descriptors from which $d'$ is reachable in the
browsing graph.
So the documents of a sample for a given descriptor with
wildcards are illustrative for several next states.
For instance, with **0 as current descriptor, a document
described by 010 can illustrate 0*0 and *10.
We then have to choose to display several times a same
miniature or not. Doing so can be acceptable if the miniatures are
cleverly organized on the screen. But simple solutions may
also work.
For instance, we can label the miniatures with the different
possibilities it illustrates as pictured in
figure~\ref{fig-interface} with $0{*}{*}$ (dark bottom) as
current descriptor.
Each document then has one label for each wildcard in the
current descriptor.
The user acts by selecting a label attached to a document,
fixing the corresponding digit with its value in that
document.


Notice that in the example of figure~\ref{fig-interface},
the sample is not minimal.
When the sample which is retrieved is not minimal, which
will generally occur with the mechanism we propose in
section~\ref{sec-sample-resolution}, some properties are illustrated several times.
We think that as long as the number of miniatures is not too
large, this is not disturbing for the user.

\section{Routing, query resolution}

\subsection{Use of a distributed hash table}
We need the classical services of distributed hash tables
(publishing, locating, dynamical set of participants...)
and we do not modify them.
As already said, we could simply reuse a DHT by choosing the
right hash function and compiling hash keys with wildcards
into sets of hash keys, but adding a routing mechanism for
wildcards ensures a better use of the network.

In this section, we show how to add that routing
mechanism into the Tapestry DHT~\cite{tapestry2001}. Our
approach is not restricted to Tapestry but the routing
mechanism must be adapted to each kind of DHT.

\subsection{Choice of the DHT}

Although our global setting is independent of the DHT, the
choice of the DHT has effects on the whole application.
One of the main advantages of Tapestry is probably that it
takes locality into account which is important in our case
because multimedia documents are relatively large files and
because browsing involves an interaction requiring low
latency.
Additionally, Tapestry hash keys are structured into digits
of several bits which can be associated to the digits of our
descriptors.

Chord~\cite{Chord2001}) does not seem to be a good candidate
compared to Tapestry since it does not have these two
properties. 
The routing system of CAN~\cite{CAN2001} could be
interesting because the digits can be resolved in a random
order, meaning that the fixed values of descriptors can be
resolved first, leaving the wildcards for the end, resulting
in less logical messages in the system. But CAN does not
take the locality into account.

\subsection{Total resolution}
\label{sec-total-resolution}

When the browsing process has ended, the nodes responsible for the descriptors
denoted by the current descriptor are contacted and they send to user's
system a list of locations for all the corresponding documents.
Then user's system can use the received physical addresses to
query the documents or their miniatures.

\newcommand{\total}{\mathit{Total}}

We add a specific routing algorithm for finding all the
nodes responsible of the denoted classes.
Let us consider a message $m= \langle d,\total,
o,i\rangle$ where 
$d$ is a descriptor with wildcards (more precisely a hash key
with wildcards),
$\total$ is a symbol meaning that a total resolution is queried,
$o$ is the physical ({\sc ip}) address of the originator of the query and
$i$ is the number of digits that have already
been resolved in the routing process.
For a node receiving this message, the behavior is the following:
\begin{description}

\item[If \mbox{$d[i]$} is not a wildcard.] The message is routed normally as
specified in the base routing mechanism of the DHT.



\item[If \mbox{$d[i]$} is a $*$ wildcard.] A new set of
descriptors with wildcards is created by replacing $d[i]$ by
each possible value in the range of the digit at
$i$ in the descriptor space.
 Each of these new descriptors $d'$ is used in a new message
$\langle d',Total, o,i\rangle$ that has to be routed.
Since $d'[i]$ is not a wildcard, each message is routed normally as
specified  in the base DHT.

\end{description}

If $n$ is the number of digits in descriptors, 
after $n$ logical hops the messages have the form $\langle
d'',Total, o,n\rangle$ (each with a different $d''$), where
the descriptors~$d''$ have no wildcards. Such a message
arrives at the node which is responsible for the descriptor
$d''$. Then, for each of these messages, the node
responsible for $d''$ sends to $o$ a list of locations
for the documents published with $d''$ as descriptor.


\subsection{Sample resolution}
\label{sec-sample-resolution}

We add a sample resolution service to the DHT to be able to retrieve
samples at low cost, that is without contacting all the
nodes responsible for all denoted classes.
Let us consider as an example the descriptor with wildcards
$d=0{*}23{*}012{*}$ (9 digits between 0 and 3), and a node
$n_0$ asking for a sample for $d$.
(We randomly consider that Tapestry resolves digits from left to
right.)
A first message is sent from $n_0$ to a neighbor node $n_1$ with a 0 as first
digit (the digit on the left) in its address, with $d$ in the message.
After this routing step, the first digit has been resolved, $n_1$ now has to resolve the second
digit, which is a $*$ wildcard.
Then the following five new descriptors are created from $d$, with a new type of wildcards used only during the routing process:
 \begin{center}
  \begin{tabular}{ccccccccc}
    0 & $\bot$ & 2 & 3 & *      & 0 & 1 & 2 & * \\
    0 & 0      & 2 & 3 & $\bot$ & 0 & 1 & 2 & $\bot$\\
    0 & 1      & 2 & 3 & $\bot$ & 0 & 1 & 2 & $\bot$\\
    0 & 2      & 2 & 3 & $\bot$ & 0 & 1 & 2 & $\bot$\\
    0 & 3      & 2 & 3 & $\bot$ & 0 & 1 & 2 & $\bot$
\end{tabular}
\end{center}
    
While we use the $*$ wildcard to mean roughly ``all'',
the bottom ($\bot$) wildcard means ``any'', that is, during the routing
process, the query can be routed to a neighbor with any value for that
digit. 

Each of these new descriptors gives a new message to be
routed. When considering a $\bot$ wildcard as the digit on
which to route, a node can replace it by the value it wishes,
in particular the best choice is to replace it by the
corresponding value in its self logical address to save a
logical hop (route to self).
Let us look at the general algorithm.

\newcommand{\sample}{\mathit{Sample}}

\subsubsection{The algorithm for retrieving samples.}

We consider the following message: $m= \langle
 d,\sample,o,i\rangle$. The behavior of a node with logical address $a$ receiving it
 is as follows:
\begin{description}

\item[If \mbox{$d[i]$} is not a wildcard.] The message is routed normally.

\item[If \mbox{$d[i]=*$.}] A new set of descriptors is created
by replacing $d[i]$ by each possible value in the range of
the digit in position $i$ in the descriptor space and all by
replacing all the other $*$ wildcards by $\bot$ wildcards.
In addition, another descriptor is created by replacing $d[i]$ by a $\bot$ wildcard (the other wildcards are not modified).
Each of these new descriptors $d'$ is used in a new message
$\langle d',\sample, o,i\rangle$ that has to be routed.

\item[If \mbox{$d[i]=\bot$}.] Any value can be chosen to instantiate $d[i]$.
Let $d'$ be $d$ where the digit in position $i$ is replaced by $a[i]$. 
This new descriptor $d'$ is used in a new message $\langle
d',\sample, o, i\rangle$ that has to be routed (the message is routed
to self, saving a logical hop).

\end{description}

\subsubsection{Do we get samples?}

This mechanism ensures that the messages reach the nodes
responsible of a (not minimal) representative set of
descriptors.
When all the digits have been resolved, each node receiving
the message sends the location of (at least) one document
having the appropriate descriptor to $o$, which is then used
to get a miniature copy of that document to be
displayed\footnote{Or the contacted machines can directly
request the nodes having miniatures to send them to $o$ to
save physical hops.}.

If for each possible descriptor there is a
document in the system having that descriptor, then this routing
mechanism ensures that the set of nodes that is finally contacted is
able to provide a correct, yet not minimal sample.

\subsubsection{Limits of the algorithm.}
A problem may occur when some descriptors are not represented.
For instance suppose the current descriptor is **0 and the sample
request reaches 010. 
If the class 010 is empty, the browsing
interface may not have a miniature to display for the 0*0 vertex
whereas 000 could contain some documents.
In such a case, an additional mechanism has to be set to
retrieve documents to represent the concerned descriptors.
This means that some documents may take more time than
others to be displayed. However:
\begin{itemize}

\item In a very large system, depending on the choice of the
descriptor function, empty classes might be rare.

\item The user does not have to wait until all the documents
are displayed to make her choice to continue the browsing
process.

\item
If a direct successor is not represented, the user can
still reach the desired class following another path.
In our example, 010 can be reached by following *10. 

\item
Some documents are representative for several direct
successors, then we have in fact several representatives for
each direct successor, then that direct successor could
still be represented by another miniature.

\item If a node in the graph does not have any illustrative
documents in the system, it seems acceptable that it should not be
accessible by browsing.

\end{itemize}

For these reasons, we think this problem does not invalidate our approach.

\subsubsection{Justification of this algorithm.}
This relatively complex algorithm to get samples is
justified by the fact that it is generally far less
expensive than getting one document for each class denoted by
an open descriptor: $O(w.b)$ nodes have to be contacted
instead of $O(b^w)$, where $w$ is the number of $*$
wildcards and $b$ is the number of possible values of a
digit (when the digits have the same range).

\section{Related work}

\paragraph{Multimedia browsing systems.}
The work of Loisant \emph{et al.}~\cite{theseLoisant,browsing2002} is a
good representative of the state of the art in multimedia
browsing systems.
It is based on a clustering of the documents that is
done before the browsing process. This pre-analysis allows to build classes which are more relevant
than the ones we use. 
However, in order to build that clustering, the system needs to
have access to all the descriptors. For this reason, that
system cannot be easily used in a peer to peer framework.
Moreover, when new documents are added to the system, the
clustering may have to be modified, which raises new difficulties.
Another difficulty in using such a system is the mapping
between clusters and logical addresses. In our system this
mapping is canonical. With dynamical or ad hoc clusters,
another DHT layer would be necessary to make the
translation.

\paragraph{Search by content in P2P systems.}
Search by content is available in some (decentralized,
structured) peer to peer systems and could have been
considered instead of using a simple DHT. 
Most of these systems are based on an indexing layers (see
the CANDy system~\cite{CANDy2004} for instance). We did not
retain this possibility because of the two DHT layers
implying a double latency. Particular features of the
systems also have to be taken into account. For instance in
CANDy a sequential process is involved to solve queries,
meaning that the more the query is complex, the longer it
takes to be resolved, which is clearly not convenient for
us.
In our system, the sub-queries are resolved in a parallel
way and the response time is the same as in the original DHT
system.

\section{Conclusion and Future work}


Our contribution is to show that structured peer to peer
systems supporting browsing can be built.
 For achieving this we use semantic hash keys instead of
 classical hash keys to provide fast access by content, with
 a minor risk of unbalance.
 We also have developed particular routing mechanisms to
 quickly retrieve sets of documents relevant to the browsing
 process.

Many points are left to the designer of the final
application. In particular, the validity of our approach
largely depends on the design of the descriptor space and of
the user interface. We have given some clues to guide that design.

A strong constraint on our system is that the structure of
the descriptors, that is the set of the properties
describing the documents, must be shared among users.
This is not convenient for open system where any kind of
document can be shared, but it is convenient for specific applications such as library catalogs or photography agency catalogs.

Further studies are necessary to validate our approach
formally, in particular the short latency and the minimal
balance of the system have to be confirmed. We could also
study how our requirements on the descriptor space limits
the possibilities of the browsing interface.

Simple features can be directly introduced in our setting : the possibility to
follow edges in the two directions in the browsing graph or the use of
intelligent features in the automatic extraction of the
descriptors (for instance, the coordinates of the center area of an image can
be dynamically computed for each image).
Some other nice features need a bit more of work, for
instance we could give several descriptors to a same
document (or use fuzzy descriptors or interval descriptors)
to cope with the fact that two very similar documents may
have different descriptors.
Finally, some features correspond to open problems, in
particular choosing the clustering of the documents
dynamically in a peer to peer way as new documents arrive
and disappear from the system (for instance, if all the
pictures are very bright, the best choice for describing
brightness is not the same as if all the pictures are very
dark or if they are well balanced in the brightness space).

\bibliographystyle{plain}

\bibliography{mabib.bib}

\end{document}